\documentclass[a4paper,onecolumn,12pt]{IEEEtran}
\usepackage{eso-pic}
\usepackage[cmex10]{amsmath}
\usepackage{amssymb,amsthm,amsfonts,mathrsfs}
\usepackage{graphicx}
\usepackage{dsfont,enumitem,stackrel}
\usepackage{color}
\usepackage{mathtools,amssymb,bm}
\usepackage{amstext}
\usepackage{setspace}
\usepackage{array}
\usepackage{algorithmicx}
\usepackage{geometry}
\usepackage[ruled]{algorithm}
\usepackage{algpseudocode}
\usepackage{algpascal}
\usepackage{algc}
\usepackage{cases}
\usepackage{lettrine}
\usepackage{pgfplots}
\usepackage{accents}
\usepackage{caption}
\usepackage{dblfloatfix}
\DeclareCaptionLabelSeparator{periodspace}{.\quad}
\usepackage{float}
\usepackage{cite}
\usepackage [autostyle, english = american]{csquotes}
\usepackage{hyperref}
\theoremstyle{remark}

\newcommand\ASTART{\bigskip\noindent\begin{minipage}[b]{0.5\linewidth}}

\newcommand\AENDSKIP{\end{minipage}\bigskip}
\newcommand\AEND{\end{minipage}}

\theoremstyle{plain}

\newtheorem{lem}{\textbf{Lemma}}

\theoremstyle{definition}

\theoremstyle{remark}

\newcommand*{\rom}[1]{\expandafter\@slowromancap\romannumeral #1@}
\graphicspath{{./Figures/}} 

\IEEEoverridecommandlockouts
\begin{document}
 
\title{Security Vulnerability of FDD Massive MIMO Systems in Downlink Training Phase}
\author{
	Mohammad~Amin~Sheikhi,~and~S.~Mohammad~Razavizadeh \\
	\IEEEauthorblockA{School of Electrical Engineering \\Iran University of Science and Technology (IUST)
		\\ee.sheikhi@gmail.com, smrazavi@ieee.org} \\
}
\maketitle
\doublespacing‎
\begin{abstract}
\textbf{We consider downlink channel training of a frequency division duplex (FDD) massive multiple-input-multiple-output (MIMO) system when a multi-antenna jammer is present in the network. The jammer intends to degrade mean square error (MSE) of the downlink channel training by designing an attack based on second-order statistics of its channel. The channels are assumed to be spatially correlated. First, a closed-form expression for the channel estimation MSE is derived and then the jammer determines the conditions under which the MSE is maximized. Numerical results demonstrate that the proposed jamming can severely increase the estimation MSE even if the optimal training signals with a large number of pilot symbols are used by the legitimate system.}       
	
\end{abstract}
\IEEEpeerreviewmaketitle

\section{Introduction}
\IEEEPARstart{M}{assive} MIMO is known as one of the main technologies in next generation of wireless networks (5G) in which the base stations (BS) in the cellular networks are supplied with a very large number of antennas \cite{Larsson2013}. This technology brings different advantages in the performance which spectral efficiency (SE) improvement is the most important one \cite{Ngo2011}. Massive MIMO can be deployed in two modes: frequency division duplex (FDD) and time division duplex (TDD). In contrast to the TDD massive MIMO systems, in the FDD mode, the SE does not always improve with the number of BS antennas and it may even degrade if the number of BS antennas gets too large. The reason is due to large overhead in downlink training of the FDD massive MIMO systems \cite{Jiang2015}. On the other hand, in the TDD mode, channel reciprocity can be utilized to estimate the downlink channel gain from the uplink training. But there are some problems in the TDD mode, for example, pilot contamination and calibration errors caused by hardware impairment can degrade the TDD massive MIMO systems performance significantly \cite{7339665},\cite{Bjornson2013}. Besides, the FDD mode has some advantages over the TDD mode, e.g. lower latency and better performance in symmetric traffic services. More importantly, most of the currently deployed systems are working in the FDD mode and economically it will be more efficient for the new generation networks to operate in the FDD mode. Therefore, using the FDD mode in massive MIMO systems has been an important research topic in recent years. One of the important problems in this regard is reducing the downlink training overhead that has been investigated in some papers, e.g. \cite{Shen2016,Gao2016,Fang2017,Choi2014,Dutta2017}. Also, there are some works on improving the efficiency of FDD massive MIMO systems with some assumptions about the channel model. For instance, the authors in \cite{Wang2016} have considered a single-user FDD massive MIMO with a correlated channel and proposed an algorithm to optimize the energy efficiency of the system by adjusting training length and transmit power.      \\
Another main issue in 5G networks is the security concerns because of their huge capacity and wider coverage. Physical layer security is one of the most effective approaches to solve the security issues of wireless networks against eavesdropping and jamming attacks \cite{8335290}. Massive MIMO has an intrinsic security against passive eavesdropping \cite{Kapetanovic2015}. But in the case of active eavesdroppers and jammers, massive MIMO security is not guaranteed and could be vulnerable. This problem has been investigated extremely in many works, e.g. \cite{Pirzadeh2016,Wu2016,Nguyen2017,8094873,8017521}.
In \cite{Pirzadeh2016}, the authors have considered a multi-user TDD massive MIMO system and demonstrated that how a limited-power smart jammer can perform an optimal attack in both uplink channel estimation and data transmission to minimize the uplink spectral efficiency of the system.
In \cite{Wu2016}, the authors have explored the pilot contamination attack by an active eavesdropper in a multi-cell TDD massive MIMO network. The secrecy rate is analyzed for matched filter precoding and an artificial random noise transmission strategy. In addition, a precoder null space design is proposed to secure the communication against the eavesdropper.  
In \cite{Nguyen2017}, the authors have studied an advanced full-duplex adversary with a massive array who tries to attack a TDD single-user massive MIMO network. The adversary simultaneously performs eavesdropping and jamming. It is shown that even with imperfect jamming channel estimation and self-interference, the jammer can still disable conventional physical layer protecting schemes.
In \cite{8094873}, the authors have proposed an approach to detect jammers in the TDD massive MIMO systems by exploiting some unused pilots in the system and showed that by increasing the number of base station antennas and unused pilots, the proposed scheme can detect the jamming more efficiently.
In \cite{8017521}, a robust jamming-resistant receiver in the uplink of a TDD massive MIMO network is designed which utilizes some purposely unused pilot symbols in the training phase.
All of the aforementioned papers and other related references therein have assumed the TDD mode for massive MIMO networks and as far as we know, no work in the literature has considered the security issues of FDD massive MIMO systems. \\
In this paper, we study the security of massive MIMO systems in FDD mode. In particular, we consider downlink channel training of an FDD massive MIMO system when there is a multi-antenna jammer in the environment who tries to attack the training phase and degrade the channel estimation performance. In contrast to many other papers in this field, we have taken into account the spatial correlation of the channels which makes the channel model more realistic. The jammer designs its attack based on the second-order statistics of its channel. We show that how a smart jammer can efficiently attack the training phase and increase the estimation error significantly. The mean square error (MSE) maximization is selected as the attack criterion and the optimal design of the jammer signal is analytically derived. Numerical results illustrate that how much the proposed attack can jeopardize the downlink training phase in this system even if the BS uses optimal pilots for channel estimation. This security vulnerability is shown to be more severe at stronger correlated channels.\\
The remainder of this paper is organized as follows. In Section II, the system model is introduced. Downlink channel training procedure is presented in Section III. In Section IV, the jamming signal design problem is formulated and solved. Numerical results are given in Section V and in the end, Section VI provides the conclusion of this paper.
\subsection{Notation}
Throughout the paper, we use boldface uppercase to denote matrices, boldface lowercase for vectors and italic letters to denote scalars. $(.)^{H}$ represents conjugate transpose and $\bm{A}(i:j)$ denotes a matrix containing columns $i$ to $j$ of a matrix $\bm{A}$. $\mathbb{E} \{ . \}$ is the expectation operator and $\bm{v} \sim CN(0,\bm{R})$ represents circularly-symmetric complex Gaussian random vectors with zero mean and covariance matrix $\bm{R}$. The $L \times L$ identity matrix is denoted by $\bm{I}_L$. For two random matrices $\bm{x}$ and $\bm{y}$, the covariance matrix is represented by $\bm{C}_{\bm{x},\bm{y}}$.

\section{System Model}\label{system.model}
We consider a single-cell network with a large-scale BS supplied with $M>>1$ antennas and a single-antenna user-equipment (UE) in the presence of a jammer who has $N$ antennas. The network operates in the FDD mode. Therefore, for downlink channel estimation, the BS transmits a training sequence to the UE, then the UE estimates the channel gain and feedbacks its estimation to the BS. The BS transmits a pilot signal, $\bm{\varphi}_m$ with the length of $L$ symbols from each of its transmit antennas. These pilots can be stacked into an $M\times L$ matrix called $\bm{\Phi}$. Unitary training sequence with the same power at each of the pilot symbols is adopted in this paper, i.e. $\bm{\Phi}^H\bm{\Phi}=\bm{I}_L$. We assume that the jammer has a prior knowledge of $L$ and transmits a jamming signal containing at least $L$ symbols from each of its antennas. The signal transmitted by the jammer can be collected into an $N\times L$ matrix called $\bm{Z}$. The received signal by the UE will be     
\begin{align}
\bm{y}=\sqrt{L P_{b}}\bm{\Phi}^{H}\bm{h}+\sqrt{L P_{j}}\bm{{Z}}^{H}\bm{g}+\bm{w},
\end{align}
where $P_{b}$ is the BS transmit power in the training phase, $\bm{h} \in \mathbb{C}^{M\times1}$ is the channel gain from the BS to the UE, $P_{j}$ is the jammer transmit power, $\bm{g} \in \mathbb{C}^{N\times1}$ is the channel gain from the jammer to the UE and $\bm{w} \sim CN(0,\sigma^2\bm{I}_L)$ models the thermal noise at the UE. \\
The channel gain from the BS to the UE is assumed to be spatially correlated. It is modeled as $\bm{h} \sim CN(0,\bm{R}_h)$ where $\bm{R}_h=\mathbb{E}(\bm{h}\bm{h}^H)$ is the covariance matrix of the channel vector $\bm{h}$. The same model is used for the channel gain from the jammer to the UE, i.e. $\bm{g} \sim CN(0,\bm{R}_g)$.
\section{Downlink channel estimation}
The UE uses the received signal in (1) to estimate $\bm{h}$ by minimum mean square error (MMSE) method \cite{Kay:1993:FSS:151045} that yields
\begin{align}
\bm{{\hat{h}}}=\bm{C}_{\bm{h},\bm{y}} \bm{C}^{-1}_{\bm{y},\bm{y}} \bm{y},
\end{align}
where the covariance matrices are computed as
\begin{align}
&\bm{C}_{\bm{h},\bm{y}}=\sqrt{L P_{b}}\bm{R}_h\bm{\Phi} \\
&\bm{C}_{\bm{y},\bm{y}}=L P_{b}\bm{\Phi}^{H}\bm{R}_h\bm{\Phi}+L P_{j}\bm{Z}^{H}\bm{R}_g\bm{Z}+\sigma^2\bm{I}_L.
\end{align}

The estimated channel gain distribution is $\bm{{\hat{h}}} \sim CN(0,\bm{\psi})$  where the covariance matrix $\bm{\psi}$ is computed as
\begin{align} \label{saigh}
\bm{\psi}=\bm{R}_h\bm{\Phi}(\bm{\Phi}^{H}\bm{R}_h\bm{\Phi}+\dfrac{P_{j}}{P_{b}} \bm{Z}^{H}\bm{R}_g\bm{Z} + \dfrac{\sigma^2}{L P_{b}}\bm{I}_L)^{-1}\bm{\Phi}^{H}\bm{R}_h.
\end{align}

We define the estimation error vector as $\bm{\epsilon}=\bm{h}-\bm{{\hat{h}}}$ that $\bm{\epsilon} \sim CN(0,\bm{R}_h-\bm{\psi})$  and the average MSE per antenna (hereafter MSE) is computed as
\begin{align}
MSE=\dfrac{1}{M}\mathbb{E}[\parallel\bm{\epsilon}\parallel_2^2].
\end{align}

By exploiting Wishart matrix properties in \cite{M2014}, the MSE will be
\begin{align}
MSE=\dfrac{1}{M}tr(\bm{R}_h-\bm{\psi}).
\end{align}

The eigenvalue decomposition (EVD) of $\bm{R}_h$ is $\bm{R}_h=\bm{U}_h\bm{D}_h\bm{U}_h^{H}$ where $\bm{D}_h=diag(\lambda_{1}^h,\lambda_{2}^h,...,\lambda_{M}^h)$ is a diagonal matrix containing the eigenvalues of $\bm{R}_h$ in descending order and $\bm{U}_h$ contains the corresponding eigenvectors. The BS does not know about the jammer presence and designs the pilot matrix $\bm{\Phi}$ to minimize the MSE without taking into account the effect of the jammer. In \cite{Choi2014}, it is shown that the optimal design of pilots to minimize the MSE is as follows
\begin{align}\label{optimalPilot}
&\bm{\Phi}_{opt}=\arg\min_{\bm{\Phi}}\dfrac{1}{M}tr(\bm{R}_h-\bm{\psi})=\bm{U}_h{(1:L)}.
\end{align}

In the next section, we will analyze the estimation performance with the above optimal pilot design in the presence of our proposed jammer signal design.
   	
\section{Jammer attack signal design}

In this section, we look at the channel estimation procedure from the jammer's point of view and show that how a smart jammer with a limited power can efficiently design its attack signal, $\bm{Z}$, to maximize the estimation error even if the BS uses the optimal pilots as in (\ref{optimalPilot}). The jammer knows its channel covariance matrix $\bm{R}_g$ since it is the second-order statistics of the channel and changes slowly over many coherence intervals. The eigenvalue decomposition (EVD) of $\bm{R}_g$ is $\bm{R}_g=\bm{U}_g{\bm{D}_g}\bm{U}_g^{H}$ where $\bm{D}_g=diag(\lambda_{1}^g,\lambda_{2}^g,...,\lambda_{N}^g)$ is a diagonal matrix containing the eigenvalues of $\bm{R}_g$ in decreasing order and $\bm{U}_g$ is corresponding eigenvector matrix. The jammer can design the signal $\bm{Z}$ in different ways. However, in all designs, the unitary signal structure with equal power at each of the symbols is used, i.e. $\bm{Z}^H\bm{Z}=\bm{I}_L$. The jammer solves the following optimization problem to design its attack signal
\begin{align} \label{zopt}
&\bm{Z}_{opt}=\arg\max_{\bm{Z}}\dfrac{1}{M}tr(\bm{R}_h-\bm{\psi}). \\
&\mathrm{s.t.}~~ \bm{Z}^H\bm{Z}=\bm{I}_L \nonumber
\end{align}
The matrix $\bm{Z}_{opt}$ that maximizes the objective function in (\ref{zopt}) minimizes $tr(\bm{\psi})$. The following lemma gives a simple equivalent problem for (\ref{zopt}) and and presents a solution for it.
\begin{lem}
An equivalent problem for (\ref{zopt}) is
\begin{align} \label{zopteq}
&\bm{Z}_{opt}=\arg\max_{\bm{Z}}tr(\bm{Z}^H\bm{R}_g\bm{Z}). \\
&\mathrm{s.t.}~~ \bm{Z}^H\bm{Z}=\bm{I}_L   \label{constraint} \nonumber
\end{align}
The $\bm{Z}_{opt}$ in (\ref{zopteq}) should satisfy these two conditions
\begin{align}
&\bm{Z}_{opt}^H\bm{R}_g\bm{Z}_{opt}=diag(\lambda_{1}^g,\lambda_{2}^g,...,\lambda_{L}^g) \\
&\bm{Z}_{opt}^H\bm{Z}_{opt}=\bm{I}_L
\end{align}
which implies that $\bm{Z}_{opt}=\bm{U}_g{(1:L)}$. \\
proof: See Appendix. 
\end{lem}
Based on this lemma, we conclude that if the BS uses $L$ symbols for downlink training, a jammer with $N \geq L$ antennas can design an optimal attack signal and maximize the MSE. In the next section, we will evaluate the performance of the proposed jamming attack by numerical simulations.

\section{Numerical Results}
In this section, the performance of the proposed jamming is explored by means of numerical simulations and we inspect the estimation MSE in different channel conditions and pilot signal designs at the BS. We consider a BS with a uniform linear array (ULA) consisting of $M=100$ antennas. The exponential correlation model is used for the covariance matrix $\bm{R}_h$ with elements $\bm{R}_{h_{i,j}}=r^{|i-j|}$, where the coefficient $r\in(0,1]$ determines the strength of the correlation in the channel \cite{Bjornson2013}. The same model is used for the jammer's channel covariance matrix. Path-loss and shadow-fading are assumed to be the same for both channels and are normalized to unity. Furthermore, the variance of thermal noise is assumed to be $\sigma^2=1$ and the transmit power of the BS and the jammer are measured in dB relative to $\sigma^2$.  \\
To show the vulnerability of the estimation procedure in the presence of the proposed jamming, we consider five different scenarios and compare them in terms of the channel estimation MSE. The BS can design the pilot signal matrix in different ways but two extreme cases are important here. In the first case, the BS uses the optimal pilots in (\ref{optimalPilot}). In the second case which is the worst case scenario, the BS uses the complementary of these pilots. We call it the worst-case pilots which are obtained by the following problem,
\begin{align}
&\bm{\Phi}_{c}=\arg\max_{\bm{\Phi}}\dfrac{1}{M}tr(\bm{R}_h-\bm{\psi})=\bm{U}_h{(M-L+1:M)}.
\end{align}
\begin{figure}[t]
	\hspace*{-.5cm}
	\centering
	\includegraphics[scale=.66]{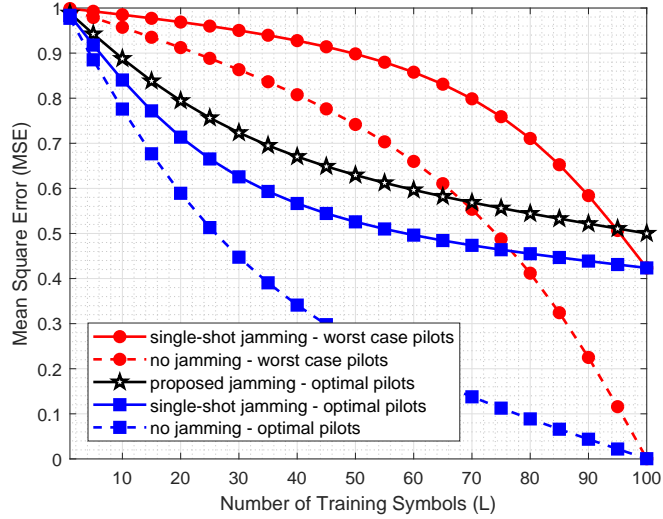}
	\caption{MSE of the system versus the number of training symbols $L$. (The channel correlation coefficient is $r=0.4$, the number of BS antennas is $M=100$ and the transmit power of the BS and the jammer are $P_{b}=P_{j}=5dB$.)} 
\end{figure}
\begin{figure}[t]
	\hspace*{-.5cm}
	\centering
	\includegraphics[scale=.66]{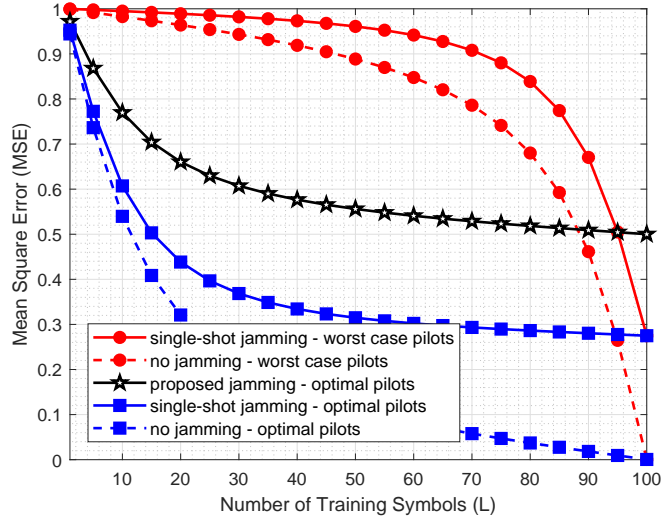}
	\caption{MSE of the system versus the number of training symbols $L$. (The channel correlation coefficient is $r=0.7$, the number of BS antennas is $M=100$ and the transmit power of the BS and the jammer are $P_{b}=P_{j}=5dB$.)} 
\end{figure}

This can be derived by following an approach similar to the proof of (\ref{optimalPilot}) in \cite{Choi2014}. In the jammer side, we consider our proposed jamming design and two other scenarios for benchmarking. First, the jammer is silent and does not attack the system. In the second scenario, the jammer designs its attack signal without considering the second-order statistics of its channel and the objective in (\ref{zopt}) and only satisfies constraint $\bm{Z}^H\bm{Z}=\bm{I}_L$. One way to do this which we call single-shot jamming is when every column of $\bm{Z}$ has only one '1' entry, and none of the rows has more than one '1' entry.
\begin{figure}[t]
	\hspace*{-.5cm}
	\centering
	\includegraphics[scale=.66]{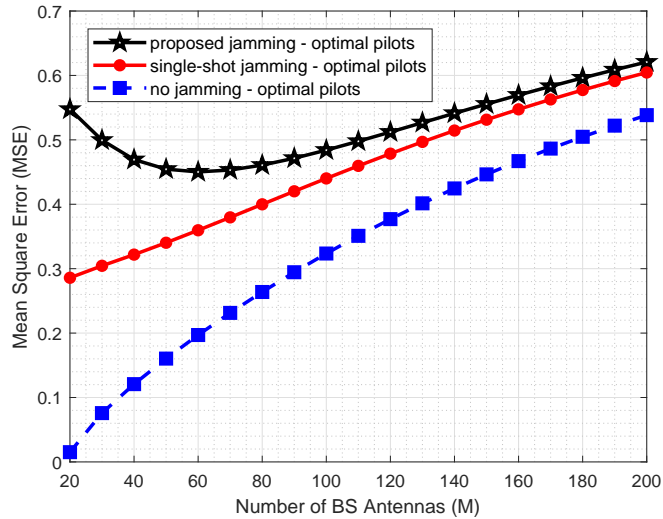}
	\caption{MSE of the downlink channel estimation versus the number of BS antennas $M$. (The channel correlation coefficient is $r=0.7$, the number of training symbols is $L=20$ and the number of jammer antennas is $N=25$.)} 
\end{figure}

Fig.~1 illustrates the MSE of the estimator versus the number of pilot symbols the aforementioned pilot and jamming signal designs. We can see that in a realistic case that the BS uses the optimal pilots $\bm{\Phi}_{opt}$, our proposed jamming has a severe effect on the MSE and makes it close to the case that the BS uses the worst-case pilots. When the number of symbols, $L$ gets close to the number of BS antennas, the MSE under the proposed jamming gets even larger than the worst-case pilots scenario. We also see that when there is no jamming in the system, the MSE will tend to zero by increasing $L$, but in the presence of the proposed jammer, it will saturate to a value around 0.5. This implies that the estimation procedure in this system is severely vulnerable to the jamming attack. The other point that can be seen from this figure is the merit of our proposed jamming in compared to single-shot jamming design. \\
Fig.~2 is in the same scenario as in Fig.~1 but with a larger correlation coefficient i.e. a stronger correlated channel. We can see that when the channel is more correlated, the optimal pilot design makes the MSE very small in the case of no jammer or with single-shot jamming in the system. But with our proposed jammer signal design, the MSE gets significantly large. Also it should be noted that in all the scenarios, when the number of pilot symbols $L$ is equal to the number of BS antennas, the MSE will be relatively small, but if the jammer uses our proposed design, the MSE will still be around 0.5 and can be very destructive in the downlink data phase precoder design. \\
Fig.~3 shows the channel estimation MSE versus the number of BS antennas. The number of pilot symbols in the system is fixed at $L=20$ and the jammer is assumed to have $N=25$ antennas. As we see, in the presence of our proposed smart jammer, the more antennas at the BS can blow down the MSE. However, after a minimum point, the MSE starts to grow up by increasing the number of BS antennas. That is because a large number of antennas leads to a high dimensional channel vector and $L=20$ pilot length is not sufficient to estimate this channel even if it is strongly correlated. Note that at any number of BS antennas, the MSE in the presence of our proposed jammer is still larger than all other scenarios that adopt optimal pilot designs at the BS.

\section{Conclusion}\label{section.conclusion}
In this work, we considered the security of an FDD massive MIMO system against a jammer who intends to attack the downlink training phase and degrade the estimation performance. The jammer tries to maximize the estimation MSE by optimal designing of its attack signal even if the BS uses the optimal training signals with a large number of pilot symbols. Numerical results showed the severe impact of this attack. In particular, when the BS uses optimal pilots with enough length of symbols, the estimation MSE could tend to zero in the absence of jammer or in the presence of other jamming schemes. But if the jammer attacks the system using our proposed design, the estimation MSE will be still large even at a large number of pilot symbols. This shows the security vulnerability in the downlink training phase of FDD massive MIMO systems against the proposed smart jammer.

\section*{Appendix}
\subsection{proof of Lemma 1}

First, we show that solving the problem in (\ref{zopteq}) is equivalent to the solution of (\ref{zopt}). As $M$ is a constant and $\bm{R}_h$ is independent of $\bm{Z}$, we have
\begin{align}
\arg\max_{\bm{Z}}\dfrac{1}{M}tr(\bm{R}_h-\bm{\psi})=\arg\min_{\bm{Z}}tr(\bm{\psi})
\end{align}

Using the fact that $tr(\bm{A}\bm{B}\bm{C})=tr(\bm{B}\bm{C}\bm{A})$, we can rewrite equation (\ref{saigh}) as follows 
\begin{align}
&tr(\bm{\psi})=tr((\dfrac{P_{j}}{P_{b}} \bm{Z}^{H}\bm{R}_g\bm{Z}+\bm{Q}_1)^{-1}\bm{Q}_2)  \\
&\bm{Q}_1=\bm{\Phi}^{H}\bm{R}_h\bm{\Phi}+\dfrac{\sigma^2}{L P_{b}}\bm{I}_L \\
&\bm{Q}_2=\bm{\Phi}^{H}\bm{R}_h^2\bm{\Phi}
\end{align}  

$\bm{Q}_1$ and $\bm{Q}_2$ are independent of $\bm{Z}$. Also note that $\bm{Z}^{H}\bm{R}_g\bm{Z}$ is in the inverted part of $\bm{\psi}$, therefore
\begin{align}
\arg\min_{\bm{Z}}tr(\bm{\psi})=\arg\max_{\bm{Z}}tr(\bm{Z}^{H}\bm{R}_g\bm{Z}).
\end{align}

To solve the equivalent problem in (\ref{zopteq}), we use the fact that for a matrix $\bm{R}_g$ and any matrix $\bm{Z}$ satisfying the constraint (\ref{constraint}), the trace of matrix $\bm{A}=\bm{Z}^{H}\bm{R}_g\bm{Z}$ is maximized when $\bm{A}$ is diagonal and also the main diagonal entries of $\bm{A}$ are maximized. By exploiting the EVD of $\bm{R}_g$ and noting that the eigenvalues of $\bm{R}_g$ are in decreasing order in $\bm{D}_g$, we conclude that the matrix $\bm{Z}_{opt}$ which maximizes $tr(\bm{A)}$ and satisfies (\ref{constraint}), must meet the following equation
\begin{align}
&\bm{Z}_{opt}^H\bm{R}_g\bm{Z}_{opt}=diag(\lambda_{1}^g,\lambda_{2}^g,...,\lambda_{L}^g)
\end{align}

which implies that $\bm{Z}_{opt}=\bm{U}_g{(1:L)}$.

\bibliographystyle{IEEE}

\begin{thebibliography}{10}
	
	\bibitem{Larsson2013}
	E.~G. Larsson, O.~Edfors, F.~Tufvesson, and T.~L. Marzetta, ``Massive mimo for
	next generation wireless systems,'' {\em IEEE Communications Magazine},
	vol.~52, pp.~186--195, February 2014.
	
	\bibitem{Ngo2011}
	H.~Q. Ngo, E.~G. Larsson, and T.~L. Marzetta, ``Energy and spectral efficiency
	of very large multiuser mimo systems,'' {\em IEEE Transactions on
		Communications}, vol.~61, pp.~1436--1449, April 2013.
	
	\bibitem{Jiang2015}
	Z.~Jiang, A.~F. Molisch, G.~Caire, and Z.~Niu, ``Achievable rates of fdd
	massive mimo systems with spatial channel correlation,'' {\em IEEE
		Transactions on Wireless Communications}, vol.~14, pp.~2868--2882, May 2015.
	
	\bibitem{7339665}
	O.~Elijah, C.~Y. Leow, T.~A. Rahman, S.~Nunoo, and S.~Z. Iliya, ``A
	comprehensive survey of pilot contamination in massive mimo 5g system,'' {\em
		IEEE Communications Surveys Tutorials}, vol.~18, pp.~905--923, Secondquarter
	2016.
	
	\bibitem{Bjornson2013}
	E.~Bj{\"o}rnson, J.~Hoydis, M.~Kountouris, and M.~Debbah, ``Massive mimo
	systems with non-ideal hardware: Energy efficiency, estimation, and capacity
	limits,'' {\em IEEE Transactions on Information Theory}, vol.~60,
	pp.~7112--7139, Nov 2014.
	
	\bibitem{Shen2016}
	W.~Shen, L.~Dai, Y.~Shi, B.~Shim, and Z.~Wang, ``Joint channel training and
	feedback for fdd massive mimo systems,'' {\em IEEE Transactions on Vehicular
		Technology}, vol.~65, pp.~8762--8767, Oct 2016.
	
	\bibitem{Gao2016}
	Z.~Gao, L.~Dai, W.~Dai, B.~Shim, and Z.~Wang, ``Structured compressive
	sensing-based spatio-temporal joint channel estimation for fdd massive
	mimo,'' {\em IEEE Transactions on Communications}, vol.~64, pp.~601--617, Feb
	2016.
	
	\bibitem{Fang2017}
	J.~Fang, X.~Li, H.~Li, and F.~Gao, ``Low-rank covariance-assisted downlink
	training and channel estimation for fdd massive mimo systems,'' {\em IEEE
		Transactions on Wireless Communications}, vol.~16, pp.~1935--1947, March
	2017.
	
	\bibitem{Choi2014}
	J.~Choi, D.~J. Love, and P.~Bidigare, ``Downlink training techniques for fdd
	massive mimo systems: Open-loop and closed-loop training with memory,'' {\em
		IEEE Journal of Selected Topics in Signal Processing}, vol.~8, pp.~802--814,
	Oct 2014.
	
	\bibitem{Dutta2017}
	B.~Dutta, R.~Budhiraja, and D.~R. Koilpillai, ``Limited-feedback low-encoding
	complexity precoder design for downlink of fdd multi-user massive mimo
	systems,'' {\em IEEE Transactions on Communications}, vol.~65,
	pp.~1956--1971, May 2017.
	
	\bibitem{Wang2016}
	Y.~Wang, C.~Li, Y.~Huang, D.~Wang, T.~Ban, and L.~Yang, ``Energy-efficient
	optimization for downlink massive mimo fdd systems with transmit-side channel
	correlation,'' {\em IEEE Transactions on Vehicular Technology}, vol.~65,
	pp.~7228--7243, Sept 2016.
	
	\bibitem{8335290}
	Y.~Wu, A.~Khisti, C.~Xiao, G.~Caire, K.~Wong, and X.~Gao, ``A survey of
	physical layer security techniques for 5g wireless networks and challenges
	ahead,'' {\em IEEE Journal on Selected Areas in Communications}, vol.~36,
	pp.~679--695, April 2018.
	
	\bibitem{Kapetanovic2015}
	D.~Kapetanovic, G.~Zheng, and F.~Rusek, ``Physical layer security for massive
	mimo: An overview on passive eavesdropping and active attacks,'' {\em IEEE
		Communications Magazine}, vol.~53, pp.~21--27, June 2015.
	
	\bibitem{Pirzadeh2016}
	H.~Pirzadeh, S.~M. Razavizadeh, and E.~Bj{\"o}rnson, ``Subverting massive mimo
	by smart jamming,'' {\em IEEE Wireless Communications Letters}, vol.~5,
	pp.~20--23, Feb 2016.
	
	\bibitem{Wu2016}
	Y.~Wu, R.~Schober, D.~W.~K. Ng, C.~Xiao, and G.~Caire, ``Secure massive mimo
	transmission with an active eavesdropper,'' {\em IEEE Transactions on
		Information Theory}, vol.~62, pp.~3880--3900, July 2016.
	
	\bibitem{Nguyen2017}
	N.~Nguyen, H.~Q. Ngo, T.~Q. Duong, H.~D. Tuan, and D.~B. da~Costa,
	``Full-duplex cyber-weapon with massive arrays,'' {\em IEEE Transactions on
		Communications}, vol.~65, pp.~5544--5558, Dec 2017.
	
	\bibitem{8094873}
	H.~Akhlaghpasand, S.~M. Razavizadeh, E.~Bj{\"o}rnson, and T.~T. Do, ``Jamming
	detection in massive mimo systems,'' {\em IEEE Wireless Communications
		Letters}, vol.~7, pp.~242--245, April 2018.
	
	\bibitem{8017521}
	T.~T. Do, E.~Bj{\"o}rnson, E.~G. Larsson, and S.~M. Razavizadeh,
	``Jamming-resistant receivers for the massive mimo uplink,'' {\em IEEE
		Transactions on Information Forensics and Security}, vol.~13, pp.~210--223,
	Jan 2018.
	
	\bibitem{Kay:1993:FSS:151045}
	S.~M. Kay, {\em Fundamentals of Statistical Signal Processing: Estimation
		Theory}.
	\newblock Upper Saddle River, NJ, USA: Prentice-Hall, Inc., 1993.
	
	\bibitem{M2014}
	A.~M. Tulino and S.~Verd\'{u}, {\em Random Matrix Theory and Wireless
		Communications}, vol.~1.
	\newblock Hanover, MA, USA: Now Publishers Inc., June 2004.
	
\end{thebibliography}

\end{document}